\documentclass[twocolumn,aps,prl,preprintnumbers]{revtex4}
\usepackage{}
\usepackage{float}
\usepackage{amsfonts}
\usepackage{bbm}
\usepackage[latin9]{inputenc}
\usepackage{amsmath}
\usepackage{amssymb}
\usepackage{graphicx}
\usepackage{mathrsfs}
\usepackage{amsfonts}
\usepackage{amsthm}
\usepackage{color}
\usepackage{txfonts}
\usepackage[colorlinks=true,citecolor=blue,linkcolor=blue,urlcolor=blue,anchorcolor=blue]{hyperref}%
\hypersetup{colorlinks=true,citecolor=blue,linkcolor=blue,urlcolor=blue}
\providecommand{\U}[1]{\protect\rule{.1in}{.1in}}
\makeatletter
\@ifundefined{textcolor}{}
{
\definecolor{BLACK}{gray}{0}
\definecolor{WHITE}{gray}{1}
\definecolor{RED}{rgb}{1,0,0}
\definecolor{GREEN}{rgb}{0,1,0}
\definecolor{BLUE}{rgb}{0,0,1}
\definecolor{CYAN}{cmyk}{1,0,0,0}
\definecolor{MAGENTA}{cmyk}{0,1,0,0}
\definecolor{YELLOW}{cmyk}{0,0,1,0}
}
\makeatother
\begin{document}
\title{Strong coupling of quantized spin waves in ferromagnetic bilayers}
\author{Zhizhi Zhang}
\author{Huanhuan Yang}
\author{Zhenyu Wang}
\author{Yunshan Cao}
\author{Peng Yan}
\email[Corresponding author: ]{yan@uestc.edu.cn}
\affiliation{School of Electronic Science and Engineering and State Key Laboratory of Electronic Thin Films and Integrated Devices, University of
Electronic Science and Technology of China, Chengdu 610054, China}

\begin{abstract}
We formulate a strong-coupling theory for perpendicular standing spin waves (PSSWs) in ferromagnetic bilayers with the interlayer exchange coupling (IEC). Employing the Hoffmann boundary condition and the energy-flow continuity across the interface, we show that the PSSWs are still quantized but with non-integral quantum numbers, {\color {red}similar to the case with a surface-pinning but in sharp contrast to that with a free surface}. The magnon-magnon coupling is characterized by the spectrum splitting which is linear with the IEC in the weak-coupling region, but getting saturated in the strong-coupling limit. Analytical predictions are verified by full micromagnetic simulations with good agreement.
\end{abstract}

\maketitle
\section{INTRODUCTION}
Hybrid system combines the advantages of different subsystems for the exploration of new phenomena and technologies \cite{Kurizki3866,Xiang_RevModPhys2013}. As a prominent quasi-particle in magnetic system, spin wave and its quantum, magnon, has attracted significant recent attention \cite{Lachance_Quirion_2019,HARDER2018,BHOI2019,Li2020HybridRev} because it can conveniently couple to superconducting resonators and qubits \cite{Huebl_PRL2013,McKenzie_PRB2019,Tabuchi405,Xie_PRA2020} that are indispensable for quantum information science and quantum computing \cite{Chumak2015}. In these studies, the coupling strength, a key parameter to characterize the interconversion between the microwave photons and magnons, has been demonstrated to be proportional to the square root of the number of spins \cite{Cao_PRB2015,Fink_PRL2009}. In practical hybrid structures, different magnetic materials possess their unique features. For example, the metallic ferromagnet like permalloy (Py, Ni$_{81}$Fe$_{19}$) has comparably higher spin density than the ferrite like yttrium iron garnet (YIG, Y$_3$Fe$_5$O$_{12}$) \cite{Hou_PRL2019,Li_PRL2019}, while YIG possesses the known lowest damping factor \cite{Kajiwara2010}.

This prerequisite has intrigued recent interests in magnon-magnon coupling between different magnonic systems , where one of them serves like a ``magnonic cavity" \cite{Xiong2020,PhysRevLett_Li2020,PhysRevLett_Chen2018,PhysRevLett_Klingler2018,Qin2018}. Compared with the magnon-photon coupling, the magnon-magnon coupling is exchange interaction dominated and localized at the interface, thus enabling the excitation of the quantized perpendicular standing spin waves (PSSWs) and short-wavelength spin waves with high group velocities \cite{Liu2018,PhysRevApplied_An2019}. {\color {red}The PSSW in a single layer manifests a set of discrete modes. Their wave vectors $n\pi/d$ ($d$ is the film thickness) and spectra, are determined not only by the dispersion relation \cite{Kalinikos_1986}, but also by the surface properties \cite{RADO_1959,RAME_1976,PUSZKARSKI_1976,Barnas_JMMM1991,Mills_PRB1992}}. {\color {red} The quantum number $n$ is a critical parameter to describe the number of the half-wavelengths contained in a PSSW mode. Intrinsically, it is supposed to be an integer under the ``natural boundary conditions" \cite{Gui_PRL_2007,Kruglyak_JPCM2014}, where the pinning of the surface spins is negligible in the single YIG or Py films because of their soft magnetism, including the weak magneto-crystalline anisotropy \cite{Yin_PRL_2006,Stancil_2009} and the weak Dzyaloshinskii-Moriya interaction (DMI) \cite{Wang_PRL_2020}. Nevertheless, an additional fraction may be introduced if local spins at the surfaces are pinned (so-called ``Rado-Weertman boundary condition" \cite{RADO_1959,Rado_PR1954,Puszkarski_2016}) by the inhomogeneity of the magnetization \cite{Tahir_PRB2015}, the shape of the magnetic structure \cite{Klos_PRB2012,Navabi_PRAppl2017} or the exchange interaction \cite{Cochran_PRB_1986,Magaraggia_PRB_2011} and the interface DMI \cite{Kostylev_JAP2014}. Besides, intensive surveys have unveiled that the PSSW spectra in the exchange coupled bilayer \cite{Vohl_PRB1989,BARNAS_JMMM1989,Stamps_PRB1994,Hillebrands_PRB1990} and multilayer structures \cite{Hillebrands_PRB1990,Krawczyk_JPCM2003,Kruglyak_JPD2017} are correlated with the interactions between the layers \cite{Vukadinovic_PRB2009,Balaz_PRB2015}, where the ``Hoffmann boundary condition" concerning the interlayer exchange coupling (IEC) is used to illustrate the interfacial spin precessions \cite{Hoffmann_1970jap,Hoffmann_1970pss,Cochran PRB1992}}.

The emerging magnon-magnon coupling allows the magnon transfer and conversion between two magnonic systems with quite distinctive properties. {\color{red} The PSSWs interacting in the two layers possess vastly different quantum numbers. Typically, it is the uniform mode in a higher saturated magnetization ($M_{\rm s}$) layer interacts with the higher mode in a lower $M_{\rm s}$ layer since they are excited at close frequencies \cite{Xiong2020,PhysRevLett_Li2020,PhysRevLett_Chen2018,PhysRevLett_Klingler2018,Qin2018}. This is different from the IEC-induced symmetric (acoustic) and antisymmetric (optic) modes in the bilayer consisting of similar sublayers \cite{Vohl_PRB1989,Hillebrands_PRB1990,Li_AFM_2016,Li_AMI_2018,Wang_APL2018,Zhang_PRL1994,Khodadadi_PRAppl2017}.} Current understanding of the magnon-magnon coupling physics relied on the macrospin approximation and the integral-quantized PSSW assumption \cite{Xiong2020,PhysRevLett_Li2020,PhysRevLett_Klingler2018}. When the thickness of magnetic film is shorter than the PSSW wavelength, macrospin model is indeed reasonable by averaging the IEC strength over the entire layer \cite{Heinrich_PRL2003,Tserkovnyak_RMP2005}. The integral-quantized PSSW approximation is valid in the limit of weak IEC, where the deviation of the wave vectors from $n\pi/d$ is negligibly small. It has been demonstrated that the frequency splitting between the hybrid modes is linear with the IEC based on the macrospin model \cite{ZHANG201879,Li_AFM_2016,Wang_APL2018,Li_AMI_2018}. Moreover, it has been shown that a phase shift was induced when spin waves propagate through a hetero-structured interface due to the mismatch of the spin-wave dispersion \cite{MAILIAN2019484,MAILIAN2018,Verba_2020}. An open question is how the rather local IEC modifies the PSSW spectra {\color {red} in both layers} and the magnon-magnon coupling.

In this work, we theoretically study the magnetization dynamics in exchange-coupled ferromagnetic bilayers. Based on the Hoffmann boundary condition and the energy flow continuity through the interface, we show that the PSSWs are quantized in an unconventional manner with non-integral quantum numbers, in contrast to that of a single layer {\color {red} with pinning-free surfaces}. The magnon-magnon coupling is linear with the IEC in the weak-coupling region, but saturated in the strong-coupling limit. Micromagnetic simulations compare well with analytical findings.

The paper is organized as follows. In Sec. \ref{Theory}, we present the analytical theory to describe the unconventional quantization of PSSWs in ferromagnetic bilayers. Numerical simulations are performed in Sec. \ref{Results} to compare with analytical results. Discussions and conclusions are drawn in Sec. \ref{Conclusion}.

\section{THEORETICAL CONSIDERATIONS}\label{Theory}
\begin{figure}[htbp!]
  \centering
  \includegraphics[width=0.4\textwidth]{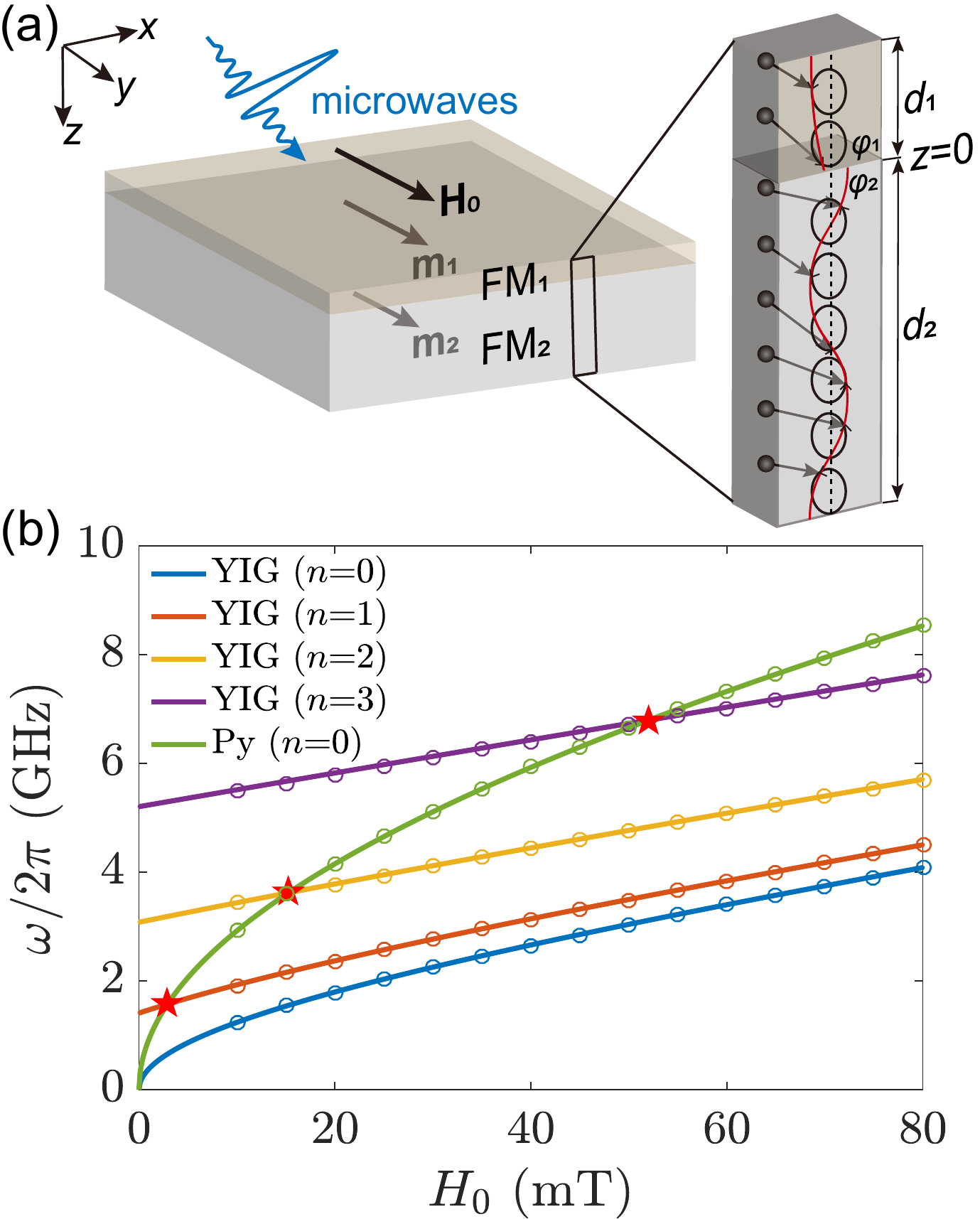}\\
  \caption{(a) Schematic of the in-plane magnetized bilayer with thicknesses of ${d_1}$ and ${d_2}$, respectively. The inset depicts the PSSW profiles across the two films. (b) Frequencies of the lowest four resonant modes of YIG (${n}=0,1,2,3$) and the Py uniform mode (${n}=0$) as a function of $ {H_0}$ when $J_{\text {int}}=0$. Solid curves are calculated based on Eq. (\ref{Disp_Rela}). Circles correspond to the micromagnetic simulations. Red pentagrams indicate the resonance of two layers.} \label{fig1}
\end{figure}

We consider a heterostructure consisting of two magnetic layers, FM$_1$ and FM$_2$ with thickness $d_1$ and $d_2$, respectively [see Fig. \ref{fig1}(a)]. The two ferromagnetic films  {\color{red} extended in $x$ and $y$ directions} are magnetized along $y$-direction and are coupled via the exchange interaction at the interface. The magnetization dynamics is governed by the Landau-Lifshitz-Gilbert (LLG) equation:
\begin{equation} \label{LLG}
\frac{\partial {\bf m}_i}{\partial t} = -\gamma \mu _0{\bf m}_i\times {\bf H}_{{\rm eff},i} + \alpha_i {\bf m}_i \times \frac{\partial {\bf m}_i}{\partial t},
\end{equation}
where $\gamma$ is the gyromagnetic ratio, $\mu _0$ is the vacuum permeability, the subscript $i=1,2$ labels each layer, $\alpha_i\ll1$ is the dimensionless Gilbert damping constant, and ${\bf m}_i={\bf M}_i/M_{{\rm s},i}$ is the unit magnetization vector with $M_{{\rm s},i}$ the saturated magnetization. The effective magnetic field ${\bf H}_{{\rm eff},i}$ comprises the intralayer exchange field $\frac{2A_{{\rm ex},i}}{\mu_0 M_{{\rm s},i}}\nabla^2 {\bf m}_i$ with the exchange constant $A_{{\rm ex},i}$, the demagnetization field ${\bf H}_{\rm d} = -m_{i,z}M_{{\rm s},i}\hat{z}$, the external field ${\bf H}_0 = H_0\hat{y}$, and the interlayer exchange field ${\bf H}_{{\rm ex,}i}$ that is given by:
\begin{equation} \label{H_eff}
\begin{aligned}
{\bf H}_{{\rm ex,}i}=J_{\text{int}}{\bf m}_j\delta(z),
\end{aligned}
\end{equation}
where $j \neq i$, $\delta(z)$ is the Dirac function, and $J_{\text{int}}=2J/\mu_0(M_{{\rm s},1}+M_{{\rm s},2})$  with the exchange energy density $J$. Neglecting the damping term and the interlayer exchange term, the dispersion relations {\color {red} of spin-waves in a two-dimensional film} can be obtained by solving the linearized LLG equation {\color {red} \cite{Herring_PR1951,Demokritov_PR2001}:
\begin{equation}\label{Disp_Rela}
\omega_i=\gamma\sqrt{\left[\frac{2A_{{\rm ex},i}}{M_{{\rm s},i}}k_{i,t}^{ 2}+\mu_0H_0\right]
\left[\frac{2A_{{\rm ex},i}}{M_{{\rm s},i}}k_{i,t}^2+\mu_0(H_0+M_{{\rm s},i}\sin\theta_i)\right]},
\end{equation}}where {\color{red} $k_{i,t}^2 = k_i^2+k_{i,\|}^2$, $k_{\|}$ is the in-plane wave vector, $k_i$} is the wave vector of the PSSW, and {\color{red} $\theta_i =\pi/2$ is the the angle between the wave vector and the static magnetization. Considering that the dynamic magnetization in $x$ and $y$ directions is uniform, it is a reasonable approximation to set $k_{i,\|}=0$.} By introducing a complex quantity $m_i^+ = m_{x,i}+im_{z,i}$, the spatial distribution of the PSSW can be expressed as:
\begin{subequations} \label{standing_wave}
\begin{align}
\label{standing_wavea} m_1^+(z) = m_{0,1}^{+}\cos(k_1z+\varphi _1), \\
\label{standing_waveb} m_2^+(z) = m_{0,2}^{+}\cos(k_2z+\varphi _2), &
\end{align}
\end{subequations}
where $m_{0,i}^{+}$ represents the precession amplitude of the PSSW, and $\varphi _i$ is the magnon phase at the interface. We assume that $\rm {FM_1}$ has a higher saturation magnetization and a thinner thickness than $\rm {FM_2}$. It is noted that the evanescent wave emerges in the $\rm {FM_1}$ layer when the frequency of the PSSW in $\rm {FM_2}$ is lower than the uniform mode in $\rm {FM_1}$ [see Fig. \ref{fig1}(b)]. In such a case, Eq. (\ref{standing_wavea}) should be modified to:
\begin{equation}\label{Anti_SW}
m_1^+(z)=m_{\rm e,1}^{+}\left [\exp\left (\frac{z}{\lambda}\right)+R\exp\left (-\frac{z}{\lambda}\right )\right ],
\end{equation}
with coefficients $m_{\rm e,1}^{+}$ and $R$ to be determined by the boundary conditions and $\lambda$ being the decay length. To determine $k_i$ in Eqs. (\ref{standing_wave}), proper boundary conditions should be implemented by integrating Eq. (\ref{LLG}) over the vicinity of the top and bottom surfaces ($z=-d_1,d_2$) and the interface ($z=0$). We assume no pinning at the two surfaces:
\begin{subequations}  \label{surface_bond}
\begin{align}
\label{surface_bonda} \frac{\partial m_{1}^{+}(z)}{\partial z}\bigg |_{z=-d_1}=0, \\
\label{surface_bondb} \frac{\partial m_{2}^{+}(z)}{\partial z}\bigg |_{z=d_2}=0.  &
\end{align}
\end{subequations}

Substituting Eqs. (\ref{standing_wave}) into Eqs. (\ref{surface_bond}), we obtain:
\begin{subequations} \label{surface_Rela}
\begin{align}
-k_1d_1+\varphi _1=n_1\pi, \\
k_2d_2+\varphi _2=n_2\pi, &
\end{align}
\end{subequations}
where $n_1$ and $n_2$ are two arbitrary integers. Instead, substituting Eq. (\ref{Anti_SW}) into Eq. (\ref{surface_bonda}), we obtain $R=\exp\left(-2d_1/\lambda\right)$, such that:
\begin{equation}\label{Anti_SWcosh}
m_1^+(z)=2m_{\rm e,1}^{+}\exp \left(-\frac{d_1}{\lambda}\right)\cosh\left(\frac{z+d_1}{\lambda}\right).
\end{equation}

At the interface ($z=0$), the combination of Hoffmann boundary conditions \cite{Barnas_JMMM1991,Hoffmann_1970jap,Hoffmann_1970pss} and the magnetic energy-flow conservation across the interface \cite{Akhiezer1968SpinWaves} leads to {\color{red}(see Appendix \ref{HoffmannBound} for detailed derivations)}:
\begin{subequations} \label{interf_bond}
\begin{align}
-\frac{2A_{{\rm ex},1}}{M_{{\rm s},1}}\frac{\partial m_{1}^{+}(z)}{\partial z}+J_{\text{int}}\left[m_{2}^{+}(z)-m_{1}^{+}(z)\right]\bigg |_{z=0}=0,\\
\frac{2A_{{\rm ex},2}}{M_{{\rm s},2}}\frac{\partial m_{2}^{+}(z)}{\partial z}+J_{\text{int}}\left[m_{1}^{+}(z)-m_{2}^{+}(z)\right]\bigg |_{z=0}=0.&
\end{align}
\end{subequations}

Substituting Eqs. (\ref{standing_wave}) into Eqs. (\ref{interf_bond}), we obtain:
\begin{subequations} \label{interface_Rela}
\begin{align}
\left(\frac{-2A_{{\rm ex},1}}{M_{{\rm s},1}}k_1\sin\varphi_1-J_{\text{int}}\cos\varphi_1\right)m_{0,1}^{+}+J_{\text{int}}\cos\varphi_2 m_{0,2}^{+}=0,\\
J_{\text{int}}\cos\varphi_1m_{0,1}^{+}+\left(\frac{2A_{{\rm ex},2}}{M_{{\rm s},2}}k_2\sin\varphi_2-J_{\text{int}}\cos\varphi_2\right)m_{0,2}^{+}=0.&
\end{align}
\end{subequations}

The resonant condition requires that the determinant of the coefficient matrix of Eqs. (\ref{interface_Rela}) vanishes:
\begin{equation} \label{interface_Cond}
\begin{aligned}
k_1k_2&\tan \left( k_1d_1\right)\tan \left(k_2d_2\right)=\\
&J_{\text{int}}\left[\frac{M_{{\rm s},1}}{2A_{{\rm ex},1}}k_2\tan \left(k_2d_2\right)+\frac{M_{{\rm s},2}}{2A_{{\rm ex},2}}k_1\tan \left(k_1d_1\right)\right].
\end{aligned}
\end{equation}

By matching the frequency $\omega_1=\omega_2$, we are able to solve the wave vectors ($k_1$ and $k_2$) and the phases ($\varphi_1$ and $\varphi_2$) in a self-consistent manner.

In this work, we focus on the coupling between the uniform mode ($k_1=0$) in $\rm {FM_1}$ and the PSSW mode ($k_2=n\pi/d_2$) in $\rm {FM_2}$ at a certain $H_0$. In the absence of the IEC, these two modes are degenerate. Magnon-magnon coupling induces a frequency splitting of the two coupled modes, which can be found by solving Eqs. (\ref{Disp_Rela}) and (\ref{interface_Cond}). It is straight-forward to see that one solution is still $k_1=0$ and $k_2=n\pi/d_2$. Since there is no explicit expression for other solutions of the transcendental equation (\ref{interface_Cond}), we adopt a graphic method. Calculation details are described in Appendix {\ref{NumerSolu}}.

Despite of the mentioned difficulty, we can obtain the perturbative solution in the limit of the weak IEC. We consider $J_{\text{int}}\rightarrow0$, so that $k_1=\Delta k_1$ and $k_2=n\pi/d_2+\Delta k_2$ with $\Delta k_1\rightarrow0$ and $\Delta k_2\rightarrow0$, and Eq. (\ref{interface_Cond}) can be expressed as:
\begin{widetext}
\begin{equation} \label{Bond_Exp}
\begin{aligned}
(\Delta k_1)^2d_1\left(\frac{n\pi}{d_2}+\Delta k_2\right)(\Delta k_2d_2)=\frac{J_{\text{int}}M_{\rm s,1}}{2A_{\rm {ex,1}}}\left(\frac{n\pi}{d_2}+\Delta k_2\right)(\Delta k_2d_2)+\frac{J_{\text{int}}M_{\rm s,2}}{2A_{\rm {ex,2}}}(\Delta k_1)^2d_1,
\end{aligned}
\end{equation}
\end{widetext}
from which, we obtain:
\begin{equation} \label{Bond_Approx}
\begin{aligned}
k_1^2\cong\frac{J_{\text{int}}M_{\rm s,1}}{2A_{\rm {ex,1}}d_1},
\end{aligned}
\end{equation}
in the weak IEC limit. {\color {red}By differentiating the dispersion relation Eq. (\ref{Disp_Rela}) with $i = 1$, we obtain:
\begin{equation} \label{Diff_Dis}
\begin{aligned}
\frac{2}{\gamma ^2}\omega\Delta\omega=\left[\frac{4A_{\rm {ex,1}}}{M_{\rm s,1}}k_1^2+\mu_0(M_{\rm s,1}+2H_0)\right]\left(\frac{4A_{\rm {ex,1}}}{M_{\rm s,1}}k_1\right)\Delta k_1.
\end{aligned}
\end{equation}

Taking into account that $\Delta k_1= k_1$, neglecting the $k_1^4$ term and putting Eq. (\ref{Bond_Approx}) into Eq. (\ref{Diff_Dis}),} we obtain the mode splitting:
\begin{equation} \label{Freq_Diff}
\begin{aligned}
\Delta\omega=\frac{\gamma^2\mu_0(M_{\rm s,1}+2H_0)}{\omega}\frac{J_{\text{int}}}{d_1}.
\end{aligned}
\end{equation}

The linear dependence of magnon-magnon coupling on $J_{\text{int}}$ is in accordance with the result of the macrospin model \cite{PhysRevLett_Li2020}. For general $J$ (or $J_{\text{int}}$), the solutions must be found numerically.

\section{NUMERICAL RESULTS AND ANALYSIS}\label{Results}
To verify our theoretical analysis, full micromagnetic simulations are performed using MUMAX3 to numerically solve the LLG equation (\ref{LLG}) \cite{Arne_2014}. The magnetic parameters of Py and YIG are used for FM$_1$ and FM$_2$, respectively. They are: $M_{{\rm s},1} = 8.6\times 10^5$ A/m, $M_{{\rm s},2} = 1.48\times 10^5$ A/m, $A_{{\rm ex},1} = 1.3\times 10^{-11}$ J/m, $A_{{\rm ex},2} = 3.1\times 10^{-12}$ J/m, $\alpha_1=0.01$, and $\alpha_2=5\times 10^{-4}$. The thicknesses are $d_1=20$ nm and $d_2=180$ nm. The $1000\times1000\times200~{\rm nm^3}$ Py/YIG bilayer structure is discretized by $50 \times 50 \times 80~{\rm }$ cells and periodic boundary conditions ($\rm {PBC}\times20$) in $x$ and $y$ directions are applied to simulate the two-dimensional infinite films. The IEC between the two layers is tuned by setting the scaling factor $S$ following Eq. (9) in Ref. \cite{Arne_2014}. {\color {red} $\alpha_i \ll 1$ has little impact on the PSSW wave vectors and the resonant frequencies, but noticeable contribution to the linewidths of the resonant peaks, which is helpful for distinguishing the YIG PSSW and Py dominant modes.} To obtain the spin-wave spectra, a microwave driving field is applied using a ``sinc" function ${\bf h}(t)=h_0 \sin [\omega _H(t-t_0)]/[\omega _H(t-t_0)] \hat x$ with the cut-off frequency $\omega _H/2\pi=50~\rm {GHz}$, $t_0 = 0.5 ~\rm {ns}$, and $h_0 = 1~\rm {mT}$. The total simulation time is 100 ns. The spin wave spectra can be analyzed from the fast Fourier transform (FFT) analysis of $m_z$.

The PSSW spectrum of an independent layer can be obtained using a ``sinc" excitation localized at the top surface of the layer {\color{red} with thickness of 2.5 nm. This excitation has a sharp square profile in the thickness direction, which contains a wide range of wave vectors covering those of PSSWs \cite{Sushruth_PRR2020}.} The resonant peaks in simulations reproduce the theoretical predictions given by Eq. (\ref{Disp_Rela}), as shown in Fig. \ref{fig1}(b). It is noted that the Py uniform mode matches with the YIG PSSW modes $n = 2 ~\rm {and}~ 3$ at $H_0=15~\rm {mT}$ and $52~\rm {mT}$ (red pentagrams), respectively. It means that the resonant energy can be transferred from one film to another with considerable efficiency if they are coupled. The uniform mode and the first-order PSSW mode frequencies of the single 20-nm thick Py layer are 6.8 and 34.3 GHz respectively at $H_0=52$ mT. This frequency gap covers several YIG layer's intrinsic PSSW modes. Below, we fix the $H_0=52~{\rm mT}$ to investigate the IEC effect on PSSW modes. We are interested in the magnon-magnon coupling near the crossing frequency $\omega_c/2\pi=6.8$ GHz. Since the IEC depends on the nature of the magnetic materials and the growth techniques, the type and strength of the coupling in the metal-oxide bilayers are not clear, particularly considering the complex crystalline and magnetic structure of YIG \cite{GELLER195730}. {\color{red} The rather local couplings might originate from the exchange interactions between the magnetic ions contained in YIG and Py, probably intermediated by some other ions (O$^{2-}$ or Y$^{3+}$) contained in the compounds. Experimental works have reported various $J$ values to be either negative \cite{PhysRevLett_Li2020,PhysRevLett_Klingler2018,Fan_PRAppl2020} or positive \cite{PhysRevLett_Chen2018,Vukadinovic_PRB2009,Chun_JAP2004} in YIG/ferromagnet bilayers.}
Therefore, we thoroughly consider the coupled bilayer structure with $J$ ranging from $-0.20$ to $1.20~ \rm {mJ/m^2}$. The negative value $J=-0.20~ \rm {mJ/m^2}$ corresponds to the antiferromagnetic exchange field $H_{\rm {ex,}1}=-42 ~\rm {mT}$ at the interface.
\begin{figure}[htbp!]
  \centering
  \includegraphics[width=0.48\textwidth]{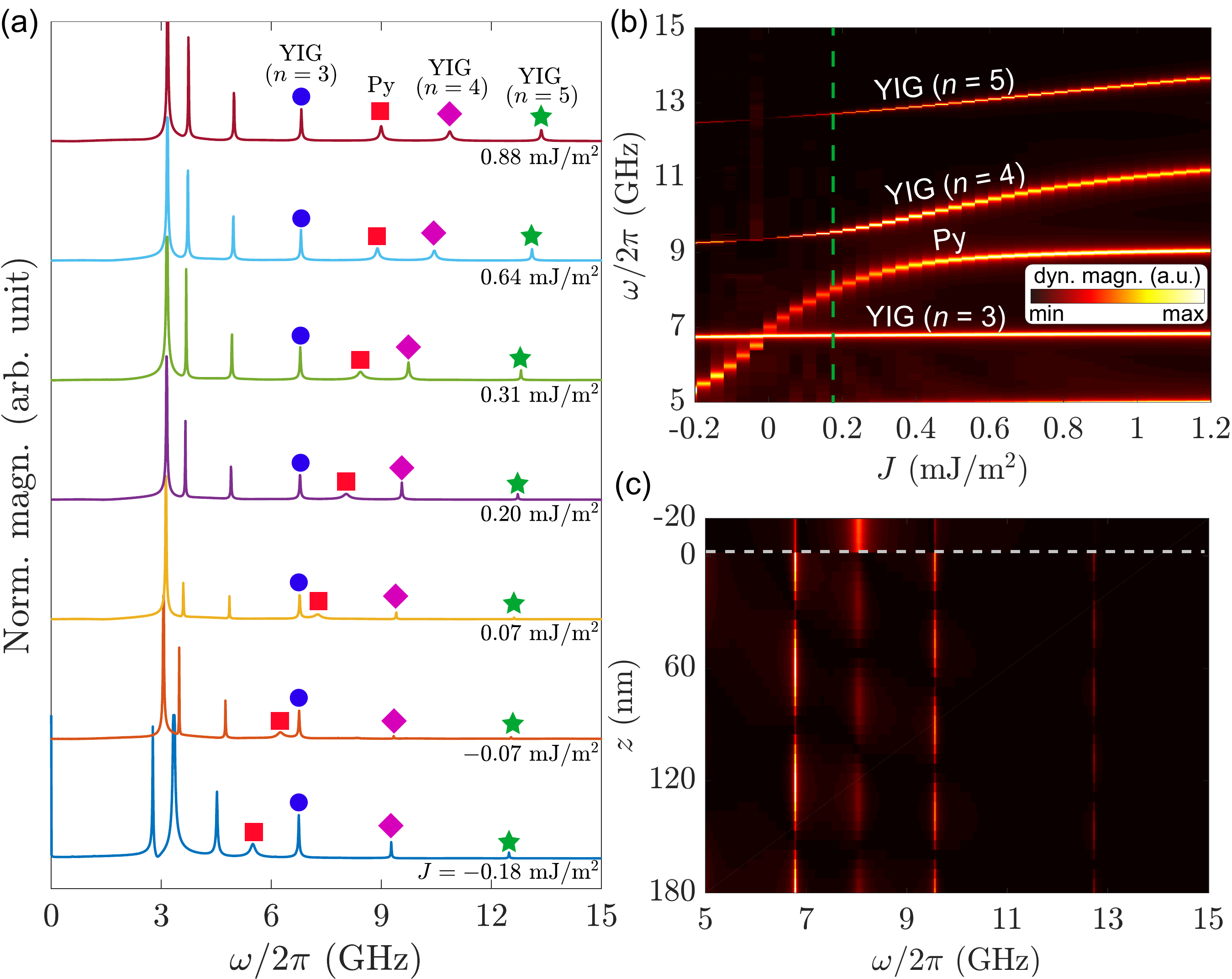}\\
  \caption{(a) The PSSW spectra in the bilayer structure with different $J$. The square, circle, diamond, and pentagram symbols represent the Py dominant mode and the YIG PSSW modes ($n$ = 3, 4, and 5), respectively. (b) The PSSW spectra as a function of $J$. (c) Spatial profiles of PSSW in the bilayer with $J=0.20~{\rm mJ/m^2}$ [labelled by the dashed green line in (b)]. The dashed white line indicates the interface.} \label{fig2}
\end{figure}

Figure \ref{fig2}(a) shows the spin wave spectra evolution in bilayer structures with representative $J$ values. The number $n$ indicates that the corresponding mode is derived primarily from the YIG PSSW mode with wave vector $n\pi/d_2$. The notation ``YIG" ( ``Py") indicates the modes with the intensive dynamic magnetization located in YIG (Py) layer. Specifically, the YIG ($n=3$) mode and the Py dominant mode can be distinguished by their linewidths near $\omega_c$ in the cases of weak coupling. All the spectra as a function of the $J$ values are summarized and color mapped in Fig. \ref{fig2}(b). Figure \ref{fig2}(c) shows the spatial distribution of the PSSW modes with $J=0.2{\rm~mJ/m^2}$. The PSSW modes in the IEC bilayer have the following features: (i) the frequency of the YIG PSSW mode ($n = 3$) remains the same with the intrinsic value $\omega _{\rm c}$, while the frequency of the Py dominant mode is lower (higher) than $\omega _{\rm c}$ for a negative (positive) $J$. (ii) The frequency of the Py dominant mode increases monotonically with the weak IEC and saturates at high IEC ($J$ is larger than $0.4~ \rm {mJ/m^2}$). (iii) The frequencies of higher-order PSSW modes ($n=4,~5,~\cdots$) increase smoothly with the IEC.
\begin{figure}[htbp!]
  \centering
  \includegraphics[width=0.48\textwidth]{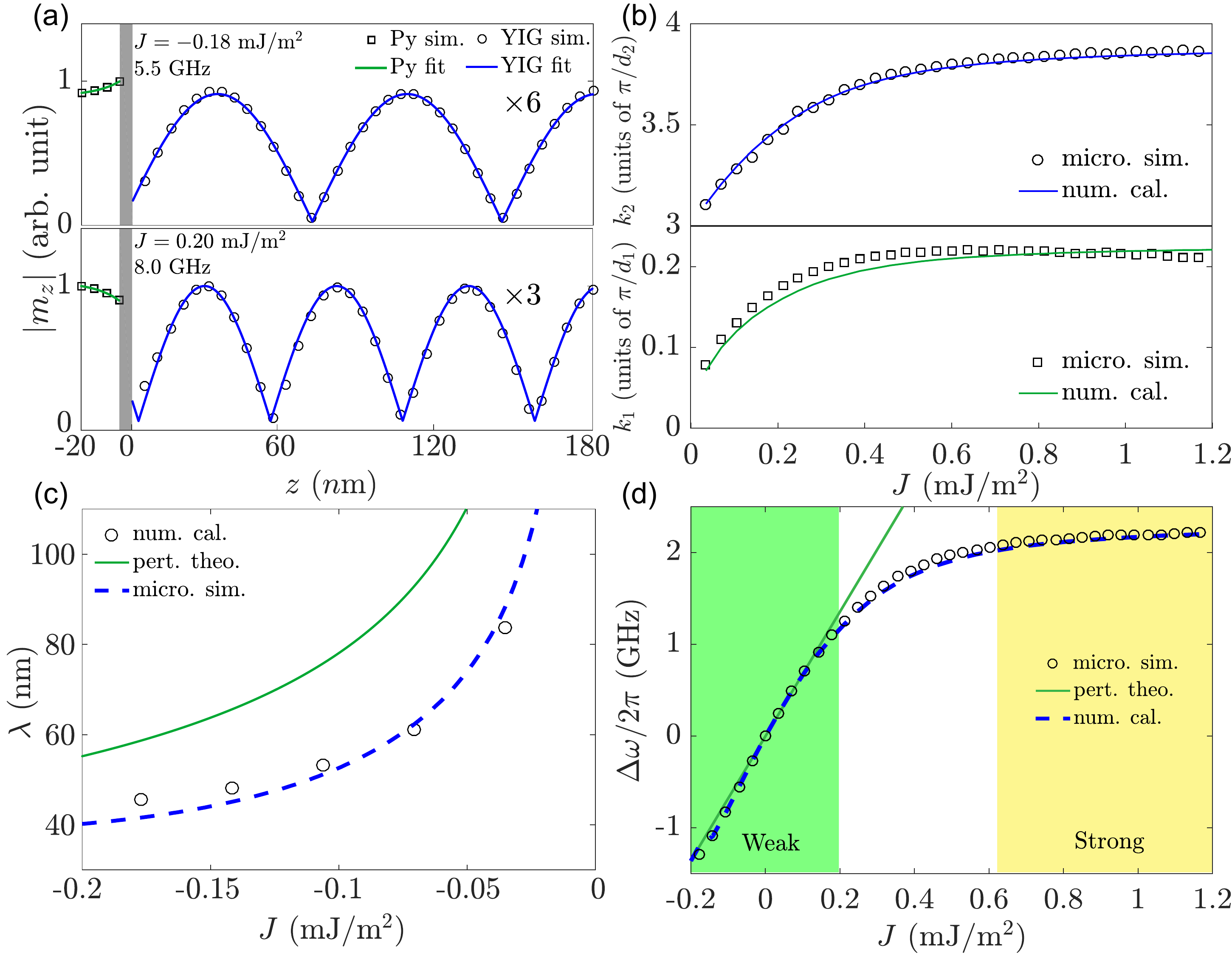}\\
  \caption{(a) Spatial distribution of $|m_z|$ of the Py dominant mode in the bilayer with the IEC strength $J = -0.18~{\rm mJ/m^2}$ at 5.5 GHz (upper panel) and $J=0.20~{\rm mJ/m^2}$ at 8.0 GHz (lower panel). The patched grey region represents the interface. Squares and circles represent the data in Py and YIG layers, respectively, obtained from micromagnetic simulations. Solid blue and green curves represent the fitting results. Since the spin-wave amplitudes in YIG are too small, we magnify them by 6 or 3 times, as labelled in each part of the figure. (b) Wave vectors in YIG (upper panel) and Py (lower panel) layers as a function of the ferromagnetic IEC strength ($J>0$). Solid curves represent the numerical solutions of Eqs. (\ref{Disp_Rela}) and (\ref{interface_Cond}) by graphic method (see Appendix {\ref{NumerSolu}}). (c) Decay length of the dynamic magnetization in the Py layer as a function of the antiferromagnetic IEC strength ($J<0$). The green line represents the analytical formula (\ref{Bond_Approx}). (d) Frequency splitting as a function of $J$. {\color {red} The patched green (yellow) part represents the linear (saturated) region.}  The solid green curves in (c) and (d) represent the formula (\ref{Freq_Diff}). The dashed blue curves in (c) and (d) represent the numerical solutions by graphic method (see Appendix {\ref{NumerSolu}}). All symbols represent the results from full micromagnetic simulations.} \label{fig3}
\end{figure}

Next, we focus on the Py dominant mode, which is highly sensitive to $J$. The intensive dynamic magnetization in Py layer excites PSSWs in YIG through the exchange torque at the interface \cite{Xiong2020,PhysRevLett_Li2020,PhysRevLett_Klingler2018,Qin2018}, which can also be regarded as the coherent spin wave source \cite{Poimanov_PRB2018}. The spatial distribution of normalized $|m_z|$ of Py dominant mode in the IEC bilayer with typical IEC values of $J=-0.18~{\rm mJ/m^2}$ at 5.5 GHz and $J=0.20~{\rm mJ/m^2}$ at 8.0 GHz are plotted in Fig. \ref{fig3}(a). The spin-wave profiles in the Py layer are sinusoidal for $J>0$ and exponentially decaying for $J<0$ (indicating evanescent spin waves as discussed in Sec. \ref{Theory}). However, spin-wave profiles in YIG layer appear to be sinusoidal in all cases. In addition, we find that the interface is not the position of the antinode, implying the introduction of a phase shift. To gain a deeper insight, we fit the spin wave profiles using appropriate functions (\ref{standing_wave}) and (\ref{Anti_SW}). To obtain the wavevectors, the spin wave profile in YIG is fitted by:
\begin{equation} \label{YIG_fit}
\begin{aligned}
\left|m_z(z)\right|=m_0\left|\cos(k_fz+\varphi _f)\right|,
\end{aligned}
\end{equation}
where $m_0$ is a scaling factor, $k_f$ and $\varphi _f$ are the fitting wave vector and phase, respectively. The sinusoidal profiles in Py shorter than half of the wavelength are accurately fitted using the second-order Taylor expansion. Assuming the maximal value located at the top surface ($z=-d_1$), in accordance with the natural boundary condition [Eq. (\ref{surface_bond})], the following formula is used for fitting the spin-wave profile in Py:
\begin{equation} \label{Py_fit}
\begin{aligned}
\left|m_z(z)\right|=m_0\left(1-\frac{\left[\left(z+d_1\right)k_f\right]^2}{2}\right).
\end{aligned}
\end{equation}

The exponentially decaying profile in Py below $\omega_c$ can be fitted by:
\begin{equation} \label{Py_exp}
\begin{aligned}
\left|m_z(z)\right|= a\exp \left(-z/\lambda_f\right)+b\exp \left(z/\lambda_f\right),
\end{aligned}
\end{equation}
where $a$ and $b$ are two scaling parameters, and $\lambda_f$ is the fitting decay length.

Figure \ref{fig3}(b) shows the wave vectors as a function of positive $J$. Both $k_1$ and $k_2$ increase rapidly near $J=0$ and saturate for large $J$. For the antiferromagnetic IEC ($J<0$), the decaying length $\lambda$ in the Py layer versus the $J$ is plotted in Fig. \ref{fig3}(c). With the increasing of the antiferromagnetic IEC strength, the spin-wave decay length decreases rapidly. Numerical solutions based on the graphic method compare very well with the full micromagnetic simulations. However, we note that the perturbation theory can only qualitatively predict the $\lambda$. The reason is that when $J\rightarrow0$, a small variation of $J$ leads to large $\lambda$ change because of the inversely-proportional relation between them, i.e., $\lambda \propto 1/ \sqrt {|J|}$, according to Eq. (\ref{Bond_Approx}). Frequency splitting ($\Delta \omega$) {\color {red}is extracted from the frequency difference between the YIG ($n = 3$) mode and the Py dominant mode. It is presented as a function of $J$} in Fig. \ref{fig3}(d). The perturbation theory [Eq. (\ref{Freq_Diff})] matches very well with both micromagnetic simulations and numerical calculations for $J$ ranging from $-0.2$ to $0.2~\rm{mJ/m^2}$ {\color {red}[green part in Fig. \ref{fig3}(d)]. It reveals that $\Delta \omega$ has an excellent linear dependence on $J$ with the slope of ${\rm 6.8~ GHz\cdot m^{2}/mJ }$ in this region. To make a comparison, the IEC strength corresponding to a bulk material can be estimated \cite{Vohl_PRB1989,Hillebrands_PRB1990} by $2A_{\rm ex}/a$ with $a$ the lattice constant (0.35 nm for Py \cite{Nahrwold_JAP2010} and 1.24 nm for YIG \cite{Stancil_2009}), considering their body-centered cubic crystalline nature \cite{Nahrwold_JAP2010,Mallmann_SSP2013}. The values of $2A_{\rm ex}/a $ for Py, YIG and their harmonic mean ($J_{12}$) are $74.3~\rm{mJ/m^2}$, $5.0~\rm{mJ/m^2}$ and $9.4~\rm{mJ/m^2}$, respectively. The linear region is within $2.1\%$ of $J_{12}$.}{\color{red} The slope of the curve rapidly decreases beyond the weak IEC range ($|J|/J_{12}>2.1\%$) and $\Delta\omega$ is saturated at $J > 0.62~\rm{mJ/m^2}$ [yellow part in Fig. \ref{fig3}(d)]. In this region, $J$ exceeds $6.6\%$ of $J_{12}$, where the slope of the curve declines to ${\rm 0.68~ GHz\cdot m^{2}/mJ }$, $10\%$ of that in linear region}. {\color {red} We have also noticed that these percentages are close to those of the frequency splitting between the symmetric and antisymmetic modes in the bilayers consisting of the same sublayers \cite{BARNAS_JMMM1989,Vohl_PRB1989,Hillebrands_PRB1990}.  }
\begin{figure*}[htbp!]
  \centering
  \includegraphics[width=0.96\textwidth]{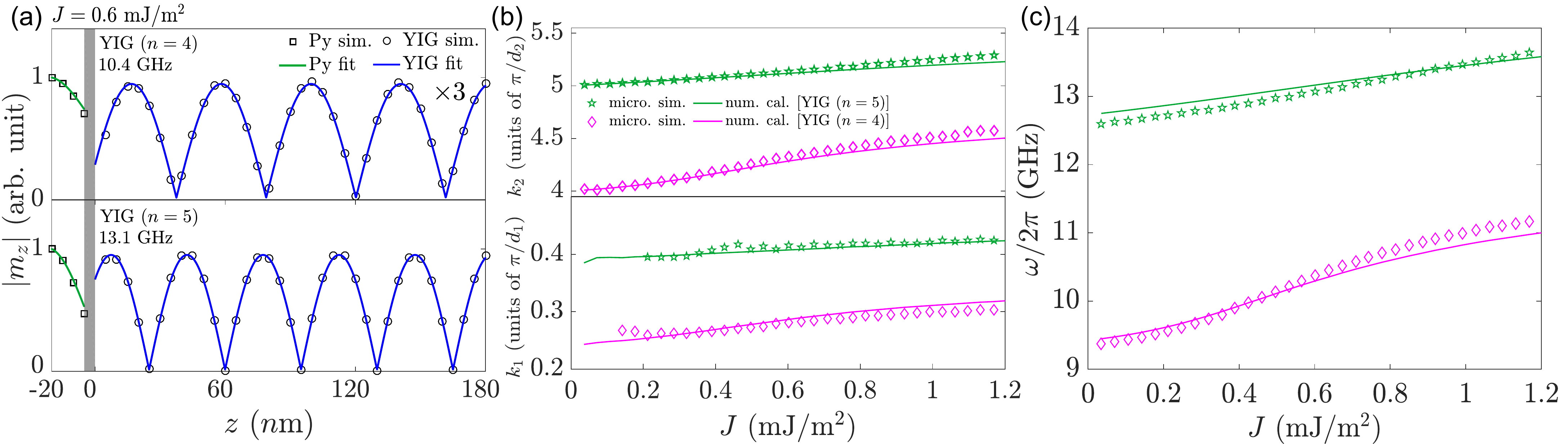}\\
  \caption{(a) Spatial distribution of $|m_z|$ for the YIG ($n$ = 4) mode at 10.4 GHz (upper panel) and the YIG ($n$ = 5) mode at 13.1 GHz (lower panel) in the bilayer with $J = 0.6~{\rm mJ/m^2}$. Since the spin-wave amplitude in YIG ($n=4$) mode is too small, we magnify them by 3 times, as labeled in the figure. (b) Wave vectors in YIG (upper panel) and Py (lower panel) layers. (c) Frequencies of the YIG ($n$ = 4 and 5) modes as a function of $J$. The magenta and green colors represent the YIG ($n$ = 4 and 5) modes respectively. Diamond and pentagram symbols correspond to the micromagnetic simulation results. Solid curves represent the numerical solutions by graphic method (see Appendix {\ref{NumerSolu}}).} \label{fig4}
\end{figure*}

It is worthy mentioning that the YIG ($n$ = 4 and 5) modes have significant frequency shifts beyond the weak IEC limit, as shown in Fig. \ref{fig2}(b). The profiles of the dynamic magnetization in both Py and YIG layers are sinusoidal, as shown in Fig. \ref{fig4}(a). Figure \ref{fig4}(b) plots the $k_1$ and $k_2$ of these modes as a function of $J$. Their frequencies can be evaluated by the dispersion relations, as plotted in Fig. \ref{fig4}(c). {\color{red} The slope of curve for the YIG ($n = 4$) mode approaches the maximum ${\rm 2.26~ GHz\cdot m^{2}/mJ }$ at $0.62~\rm{mJ/m^2}$, and that for the YIG ($n = 5$) mode maintains at ${\rm 0.91~ GHz\cdot m^{2}/mJ }$ in the investigated $J$ range. These slopes, reflecting the sensitivities of higher-order PSSW modes to $J$, are larger than that of $\Delta \omega$ in the strong coupling range.} Micromagnetic simulations (symbols) agree well with the numerical calculations (curves). {\color {red}These behaviors provide the possibility for the determination of $J$ from the PSSW spectra in the strong coupling region \cite{Vohl_PRB1989}.}

Physically, the IEC torque exerted on the spins is proportional to $J\textbf{m}_1(0) \times \textbf{m}_2(0)$, favoring the the uniform precession with positive $J$ at the interface. In the presence of a strong IEC, $\textbf{m}_1(0)$ becomes parallel with $\textbf{m}_2(0)$ \cite{Vohl_PRB1989,VAYHINGER1988307}. An even stronger IEC strength does not facilitate more interaction between the quantized magnons in the two layers. This scenario explains the saturation of $\Delta\omega$ and the frequency shifts of the higher modes with $J$. In such case, $\varphi_1$ is equal to $\varphi_2$ in Eq. (\ref{surface_Rela}), leading to $(k_1d_1+k_2d_2)/\pi=n, n\in\mathbb{N}$. Generally, the quantum number $k_id_i/\pi$ indicates how many half-wavelengths are contained within the layer. It reveals that the sum $(k_1d_1+k_2d_2)/\pi$ in the coupled bilayer becomes an integer in the strong IEC region, even though the number $k_id_i/\pi$ does not.
\begin{figure}[htbp!]
  \centering
  \includegraphics[width=0.36\textwidth]{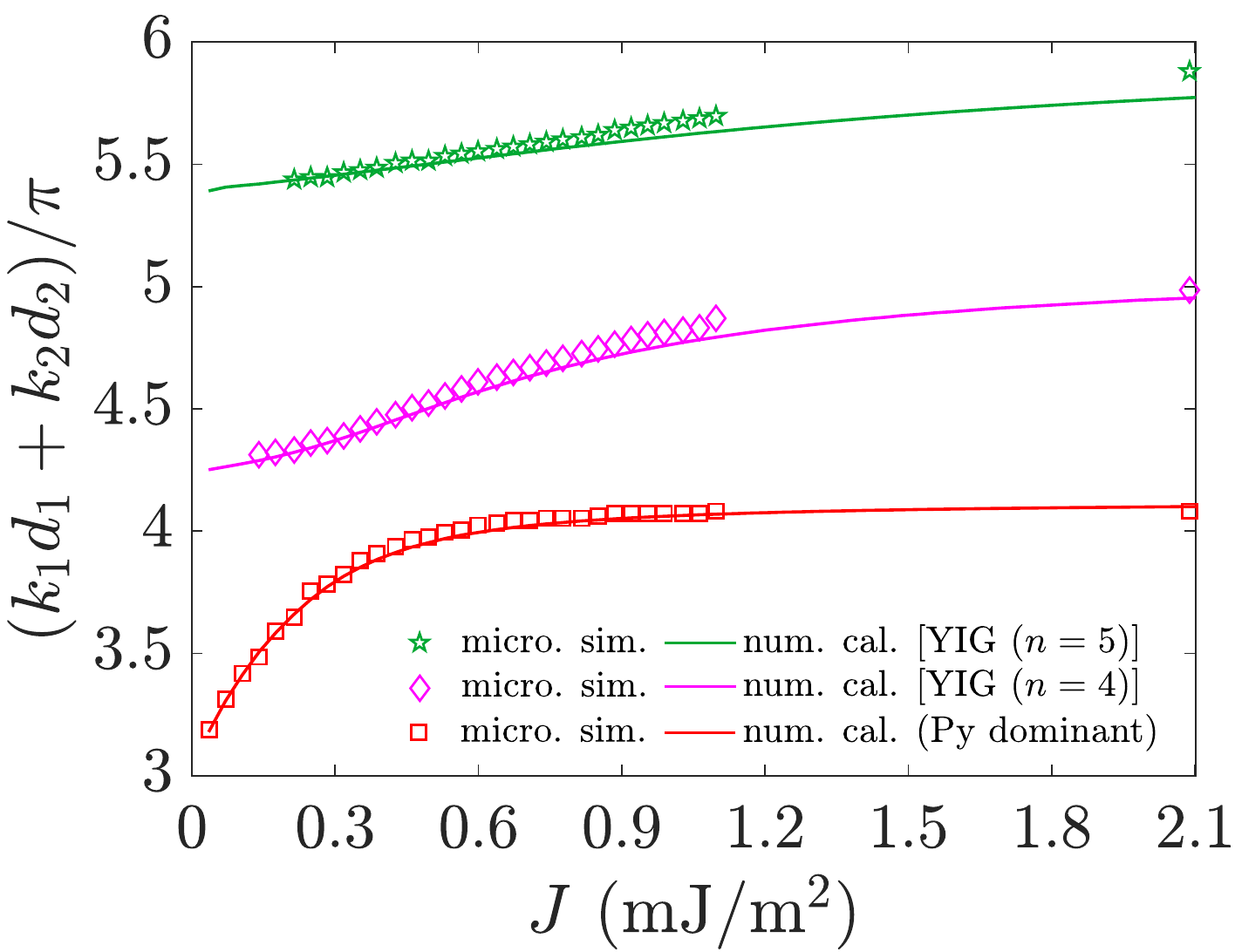}\\
  \caption{ Sum of the wave number $(k_1d_1+k_2d_2)/\pi$ in the coupled bilayer as a function of $J$. Symbols are micromagnetic simulation results. Solid curves represent the data based on numerical solutions by graphic method (see Appendix {\ref{NumerSolu}}).} \label{fig5}
\end{figure}
To verify this picture, we extend the theoretical analysis to $J=2.1\rm ~{mJ/m^2}$ and perform additional micromagnetic simulations. The dependence of $(k_1d_1+k_2d_2)/\pi$ on $J$ is plotted in Fig. \ref{fig5}. Taking the YIG ($n = 4$) mode as an example, the intrinsic mode is excited around 9.4 GHz. At this frequency, $k_1d_1/\pi$ is equal to 0.24 according to the dispersion relation [Eq. (\ref{Disp_Rela})]. It increases to 0.35 with $J\rightarrow\infty$. Accordingly, $k_2d_2/\pi$ varies from 4 to 4.65 [Fig. \ref{fig4}(b)]. For this mode, $(k_1d_1+k_2d_2)/\pi$ approaches to the next integral number 5 at $J=2.1\rm~{mJ/m^2}$, indicated as the magenta curves and symbols in Fig. \ref{fig5}. Meanwhile, $(k_1d_1+k_2d_2)/\pi$ for YIG ($n=5$) mode keeps increasing in this wide $J$ range. It suggests that the precessions of $\textbf{m}_1(0)$ and $\textbf{m}_2(0)$ in this mode remain unparallel.

\section{DISCUSSION AND CONCLUSION}\label{Conclusion}
In the above analysis, the parameters of YIG and Py are used for analysis. We point out that the proposed theory is applicable for general IEC bilayers. Experimentally, the known strongest IEC reported on the garnet/metal heterostructure is $0.86 \rm~{mJ/m^2}$, which was grown on the silicon substrate \cite{Fan_PRAppl2020}. Even stronger IEC can be achieved with the improvement of the interface quality \cite{Morrison_JAP2015}. Our results open a new perspective for the precise measure of $J$ {\color {red}from the spin wave spectra. In the weak coupling region, the frequency splitting between the coupled modes can be used to determine $J$ with its linear dependence and the large slope. Beyond this region, even though the frequency splitting is gradually saturated, we can still figure out the value of $J$ according to the frequency shifts of the higher-order PSSW modes for their higher sensitivities in this range from the spectra. These findings provide the potential to expand the range for the measure of $J$ \cite{Vohl_PRB1989} and assist with the research on tuning IEC \cite{Li_AMI_2018}}.

In summary, we investigated the IEC effects on the PSSW modes in ferromagnetic bilayers. In sharp contrast to common wisdom, we showed that the magnon-magnon coupling characterized by the frequency splitting is linear with IEC in the weak coupling limit but saturated in the strong coupling region. The frequency shifts of higher-order PSSW modes possess higher sensitivity to the IEC than the fundamental mode in the strong coupling region. Our findings are helpful for analytically understanding the strong magnon-magnon coupling in magnetic heterostructures and for precisely measuring $J$ in a wide range.

\section{ACKNOWLEDGEMENTS}
\begin{acknowledgments}\label{Acknowledgments}
We thank Z.-X. Li for helpful discussions. This work was supported by the National Natural Science Foundation of China (Grants No. 12074057, No. 11604041, and No. 11704060). Z.Z. acknowledges the financial support of the China Postdoctoral Science Foundation under Grant No. 2020M673180. Z.W. acknowledges the financial support of the China Postdoctoral Science Foundation under Grant No. 2019M653063.
\end{acknowledgments}

\begin{appendix}
{\color{red}
\section{DERIVATION OF HOFFMANN BOUNDARY CONDITION}\label{HoffmannBound}
We start from the atomistic spin Hamiltonian. Assuming that the spins are aligned in $z$ direction, normal to the interface and introducing the complex quantity $S_n^+ = S_x + iS_z$, the quantum mechanical equation describing the $n$th spin is:
\begin{equation} \label{QuanSpinEnergy}
\begin{aligned}
(E-E_n)S_n^++S_nJ_{n,n+1}S_{n+1}^++S_nJ_{n,n-1}S_{n-1}^+=0,
\end{aligned}
\end{equation}
where $E$ is the energy accumulated in the single spin, $J_{n,n+1}$ denotes the exchange constant between the $n$th and $(n+1)$th spins, and $E_n$ is defined as:
\begin{equation} \label{QuanSpinO}
\begin{aligned}
E_n=g_n{\mu_0}{\mu_B}H+S_{n+1}J_{n,n+1}+S_{n-1}J_{n,n-1},
\end{aligned}
\end{equation}
where $g_n$ is the Land\'{e} factor, ${\mu_0}$ is the vacuum permeability, ${\mu_{\rm B}}$ is the Bohr magneton, and $H$ is the external magnetic field. Since $H$ is along $y$ direction and $S_n^+$ is in $xz$ plane, $g_n{\mu_0}{\mu_B}HS_n^+\equiv 0$. The energy flow conservation requires that $E = 0$. Eq. (\ref{QuanSpinEnergy}) can be rewritten as:
\begin{equation} \label{SpinEnerCons}
\begin{aligned}
-S_{n+1}J_{n,n+1}S_n^+&-S_{n-1}J_{n,n-1}S_n^+\\
&+S_nJ_{n,n+1}S_{n+1}^++S_nJ_{n,n-1}S_{n-1}^+=0.
\end{aligned}
\end{equation}

We assume the bilayer structure with interface located at the first spin, which leads to $S_n = S_A (S_B)$ for $n < 0~ (n \geq 0)$, $J_{n,n+1} = J_A (J_B)$ for $n < -1 ~(n \geq 0)$, $J_{-1,0} = J_{AB}$ with $J_{AB}$ the exchange constant between the materials FM$_1$ and FM$_2$, and finally $g_n=g_A=g_B=g$. The equations describing the spins at the interface are:
\begin{subequations}  \label{InterfSpin}
\begin{align}
\label{InterfSpina} -S_BJ_{AB}S_{A,-1}^+-S_AJ_{A}S_{A,-1}^++S_AJ_{AB}S_{B,0}^++S_AJ_{A}S_{A,-2}^+&=0, \\
\label{InterfSpinb} -S_BJ_{B}S_{B,0}^+-S_AJ_{AB}S_{B,0}^++S_BJ_{B}S_{B,1}^++S_BJ_{AB}S_{A,-1}^+&=0.
\end{align}
\end{subequations}

The relation between $m_{1(2)}^+(z)$ and $S^+$ is given by:
\begin{equation} \label{Disc_to_Contin}
\begin{aligned}
m_{1(2)}^+(z)\bigg |_{z=0}=\frac{S_{A(B),-1(0)}^+}{S_{A(B)}}.
\end{aligned}
\end{equation}

Assuming $S^+$ as a continuous variable, it can be expanded in the first order approximation:
\begin{subequations}  \label{SpinTaylor}
\begin{align}
\label{SpinTaylora} S_{A,-2}^+=&S_{A,-1}^+-S_Aa\frac{\partial}{\partial z}m_{1}(z)^+\bigg |_{z=0}, \\
\label{SpinTaylorb} S_{B,1}^+=&S_{B,0}^++S_Ba\frac{\partial}{\partial z}m_{2}(z)^+\bigg |_{z=0},
\end{align}
\end{subequations}
where $a$ is the mesh size. Substituting Eqs. (\ref{Disc_to_Contin}) and (\ref{SpinTaylor}) into Eqs. (\ref{InterfSpin}), we obtain:
\begin{subequations}  \label{BoundSpin}
\begin{align}
\label{BoundSpina} J_{AB}S_AS_B\left [-m_1^+(z)+m_2^+(z)\right ]-J_{A}S_A^2a\frac{\partial}{\partial z}m_1^+(z)\bigg |_{z=0}&=0, \\
\label{BoundSpinb}  J_{AB}S_AS_B\left [m_1^+(z)-m_2^+(z)\right ]+J_{B}S_B^2a\frac{\partial}{\partial z}m_2^+(z)\bigg |_{z=0}&=0.
\end{align}
\end{subequations}
\begin{figure}[htbp!]
  \centering
  \includegraphics[width=0.48\textwidth]{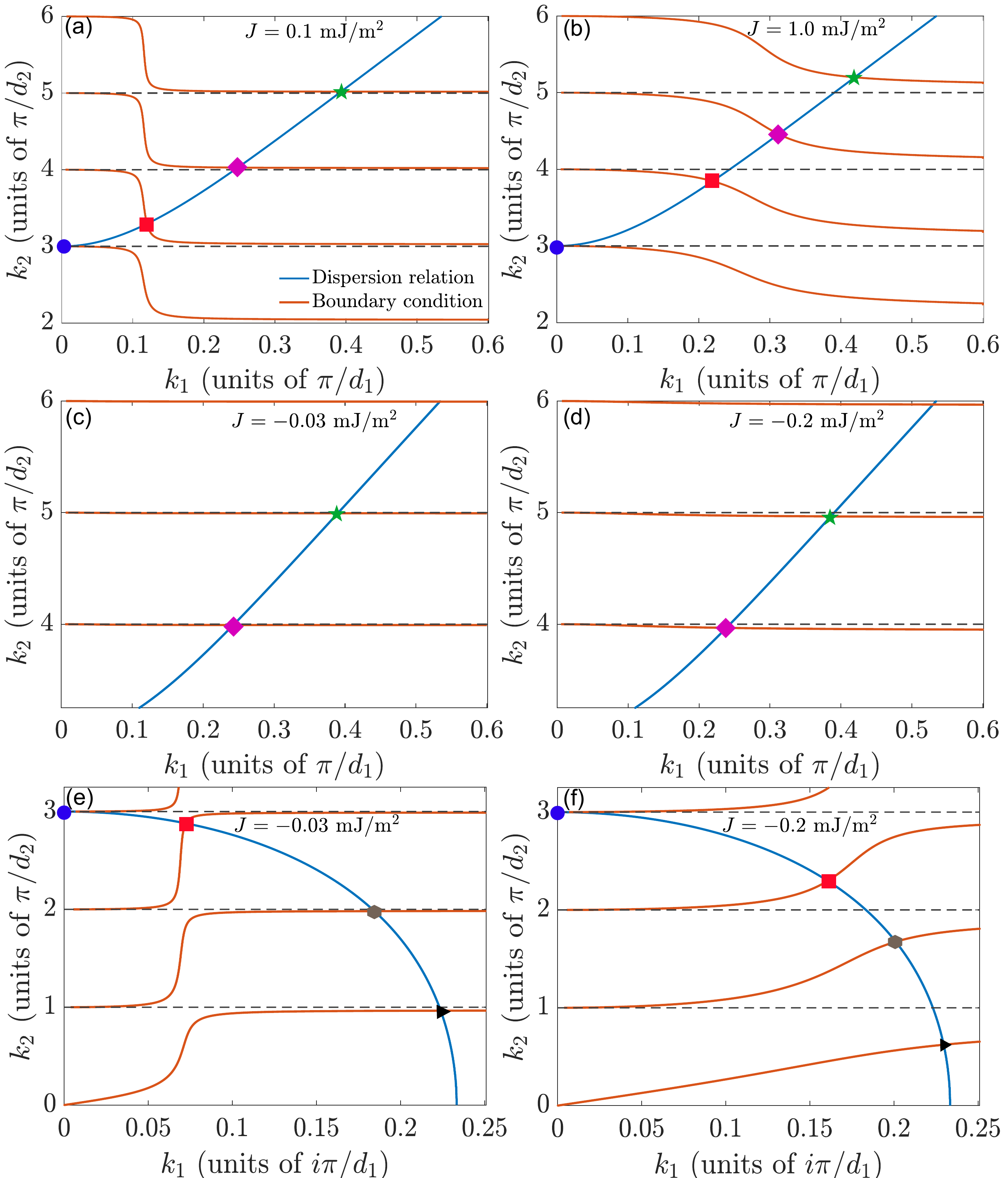}\\
  \caption{Dispersion relations (blue curves) [Eq. (\ref{Disp_Rela}) with $\omega_1 = \omega_2$] and boundary conditions (orange curves) [Eq. (\ref{interface_Cond})] to graphically determine $k_1$ and $k_2$: with $J=0.1{\rm~mJ/m^2}$ (a), $J=1.0{\rm~ mJ/m^2}$ (b), $J=-0.03{\rm~ mJ/m^2}$ (c), and $J=-0.2{\rm~ mJ/m^2}$ (d), where $k_1$ is a real number; and with $J=-0.03{\rm~ mJ/m^2}$ (e) and $J=-0.2{\rm~ mJ/m^2}$ (f), where $k_1$ is purely imaginary number. The square, triangle, hexagon, circle, rhombus, and pentagram symbols represent solutions of the Py dominant mode and the YIG PSSW modes ($n$ = 1, 2, 3, 4 and 5), respectively.}\label{fig6}
\end{figure}
Here, we have the relations between the microscopic $J_{A(B)} ~(J_{AB})$ and the macroscopic $A_{\rm ex,1(2)} ~(J)$
\begin{subequations} \label{Bulk_Exch}
\begin{align}
\label{Bulk_Excha} J_{A}&=\frac {2A_{\rm ex,1} g \mu_{\rm B}}{S_{A}^2M_{\rm s,1}a^2},\\
\label{Bulk_Exchb} J_{B}&=\frac{2A_{\rm ex,2}g\mu_{\rm B}}{S_{B}^2M_{\rm s,2}a^2},\\
\label{Bulk_Exchc} J_{AB}&=\frac{2Jg\mu_{\rm B}}{S_{A}S_{B}M_{s,\rm int}a},
\end{align}
\end{subequations}
where $M_{\rm s, int}=(M_{\rm s,1}+M_{\rm s,2})/2$ is the $M_{\rm s}$ in the interfacial region. We note that a similar relation obtained in Ref. \cite{Barnas_JMMM1991} differs with ours by the factor $g\mu_{\rm B}/(M_{s, i} a^3)~(i=1,2 {\rm ~and ~int})$, caused by the ratio between the lattice constant and the mesh size.

Substituting Eqs. (\ref{Bulk_Exch}) into Eqs. (\ref{BoundSpin}), we obtain:
\begin{subequations} \label{Hoff_Eq}
\begin{align}
-\frac{2A_{{\rm ex},1}}{M_{{\rm s},1}}\frac{\partial m_{1}^{+}(z)}{\partial z}+J_{\text{int}}\left[m_{2}^{+}(z)-m_{1}^{+}(z)\right]\bigg |_{z=0}=0,\\
\frac{2A_{{\rm ex},2}}{M_{{\rm s},2}}\frac{\partial m_{2}^{+}(z)}{\partial z}+J_{\text{int}}\left[m_{1}^{+}(z)-m_{2}^{+}(z)\right]\bigg |_{z=0}=0,&
\end{align}
\end{subequations}
which is Eqs. (\ref{interf_bond}) in the main text.}

\section{GRAPHIC METHOD TO SOLVE THE TRANSCENDENTAL EQUATION}\label{NumerSolu}
Using the parameters in Sec. \ref{Results}, the dispersion relations are plotted as the blue curves in Fig. \ref{fig6}. It is noted that when $k_2\geq3\pi/d_2$ ($k_2<3\pi/d_2$), the identity $\omega_1(k_1)=\omega_2(k_2)$ can be established with $k_1$ to be a real (purely imaginary) number, as shown by symbols in Figs. \ref{fig6}(a)-\ref{fig6}(d) [Figs. \ref{fig6}(e) and \ref{fig6}(f)].

The boundary condition curves are plotted as the orange curves in Figs. \ref{fig6}(a)-\ref{fig6}(f). The intersections between the blue and orange curves are the solutions for the PSSW modes. The blue circle, corresponding to the YIG ($n=3$) mode, is consistently located at $(k_1,k_2)=(0,3\pi/d_2)$. In the ferromagnetic IEC cases ($J>0$) [Figs. \ref{fig6}(a) and \ref{fig6}(b)], the red square symbol, corresponding to the Py dominant mode, moves rapidly at a lower $J$ value, approaching to the asymptotic line $k_2=4\pi/d_2$. It is noted that the red square and the blue circle coincide at  $(k_1,k_2)=(0,3\pi/d_2)$ with $J=0$. The purple rhombus and the green pentagram, corresponding to the YIG ($n$ = 4 and 5) modes respectively, are shifted from $k_2=n\pi/d_2$ toward $k_2=(n+1)\pi/d_2$ ($n$ = 4 and 5) with the increase of $J$. The boundary condition curve becomes flat when $J$ exceeds $2.0~\rm {mJ/m^2}$ (see Supplemental movie CurveEvolution1.gif). In this case, the coordinations ($k_1,k_2$) of the purple rhombus (green pentagram) satisfy that $k_1d_1+k_2d_2\cong5\pi$ ($k_1d_1+k_2d_2\cong6\pi$). Meanwhile, in the antiferromagnetic IEC ($J<0$) cases, the purple rhombus (green pentagram) moves little away from the horizontal line $k_2=4\pi/d_2$ ($k_2=5\pi/d_2$) with $k_1$ a real number, as shown in Figs. \ref{fig6}(c) and \ref{fig6}(d). The red square is found with $k_1$ a purely imaginary number, as shown in Figs. \ref{fig6}(e) and \ref{fig6}(f). If the antiferromagnetic IEC strength increases, the red square approaches to the horizontal line $k_2=2\pi/d_2$ (see Supplemental movie CurveEvolution2.gif).

\end{appendix}


\begin{thebibliography}{99}

\bibitem {Kurizki3866} G. Kurizki, P. Bertet, Y. Kubo, K. M{\o}lmer, D. Petrosyan, P. Rabl, and J. Schmiedmayer, Quantum technologies with hybrid systems, \href{https://www.pnas.org/content/112/13/3866}{Proc. Natl. Acad. Sci. U.S.A. \textbf{112}, 3866 (2015)}.

\bibitem {Xiang_RevModPhys2013}Z.-L. Xiang, S. Ashhab, J. Q. You, and F. Nori, Hybrid quantum circuits: Superconducting circuits interacting with other quantum systems, \href{https://link.aps.org/doi/10.1103/RevModPhys.85.623}{Rev. Mod. Phys. \textbf{85}, 623 (2013)}.

\bibitem {Lachance_Quirion_2019}D. Lachance-Quirion, Y. Tabuchi, A. Gloppe, K. Usami, and Y. Nakamura, Hybrid quantum systems based on magnonics, \href{https://doi.org/10.7567/1882-0786/ab248d}{Appl. Phys. Express \textbf{12}, 070101 (2019)}.

\bibitem {HARDER2018}M. Harder and C.-M. Hu, Cavity spintronics: An early review of recent progress in the study of magnon-photon level repulsion, \href{https://doi.org/10.1016/bs.ssp.2018.08.001}{Solid State Phys. \textbf{69}, 47 (2018)}.

\bibitem {BHOI2019}B. Bhoi and S.-K. Kim, Photon-magnon coupling: historical perspective, status, and future directions, \href{https://doi.org/10.1016/bs.ssp.2019.09.001} {Solid State Phys. \textbf{70}, 1 (2018)}.

\bibitem {Li2020HybridRev}Y. Li, W. Zhang, V. Tyberkevych, W.-K. Kwok, A. Hoffmann, and V. Novosad,  Hybrid magnonics: Physics, circuits, and applications for coherent information processing, \href{https://doi.org/10.1063/5.0020277}{J. Appl. Phys. \textbf{128}, 130902 (2020)}.

\bibitem {Huebl_PRL2013}H. Huebl, C. W. Zollitsch, J. Lotze, F. Hocke, M. Greifenstein, A. Marx, R. Gross, and S. T. B. Goennenwein, High Cooperativity in Coupled Microwave Resonator Ferrimagnetic Insulator Hybrids, \href{https://link.aps.org/doi/10.1103/PhysRevLett.111.127003}{Phys. Rev. Lett. \textbf{111}, 127003 (2013)}.

\bibitem {McKenzie_PRB2019}L. McKenzie-Sell, J. Xie, C.-M. Lee, J. W. A. Robinson, C. Ciccarelli, and J. A. Haigh, Low-impedance superconducting microwave resonators for strong coupling to small magnetic mode volumes, \href{https://link.aps.org/doi/10.1103/PhysRevB.99.140414}{Phys. Rev. B \textbf{99}, 140414(R) (2019)}.

\bibitem {Tabuchi405}Y. Tabuchi, S. Ishino, A. Noguchi, T. Ishikawa, R. Yamazaki, K. Usami, and Y. Nakamura, Coherent coupling between a ferromagnetic magnon and a superconducting qubit, \href{http://dx.doi.org/10.1126/science.aaa3693}{Science \textbf{349}, 405 (2015)}.

\bibitem {Xie_PRA2020}J.-K. Xie, S.-L. Ma, and F.-L. Li, Quantum-interference-enhanced magnon blockade in an yttrium-iron-garnet sphere coupled to superconducting circuits, \href{https://link.aps.org/doi/10.1103/PhysRevA.101.042331}{Phys. Rev. A  \textbf{101}, 042331 (2020)}.

\bibitem {Chumak2015}A. V. Chumak, V. I. Vasyuchka, A. A. Serga, and B. Hillebrands, Magnon spintronics, \href{https://doi.org/10.1038/nphys3347}{Nat. Phys. \textbf{11}, 453 (2015)}.

\bibitem {Cao_PRB2015}Y. Cao, P. Yan, H. Huebl, S. T. B. Goennenwein, and G. E. W. Bauer, Exchange magnon-polaritons in microwave cavities, \href{https://link.aps.org/doi/10.1103/PhysRevB.91.094423}{Phys. Rev. B \textbf{91}, 094423 (2015)}.

\bibitem {Fink_PRL2009}J. M. Fink, R. Bianchetti, M. Baur, M. G\"{o}ppl, L. Steffen, S. Filipp, P. J. Leek, A. Blais, and A. Wallraff, Dressed Collective Qubit States and the Tavis-Cummings Model in Circuit QED, \href{https://link.aps.org/doi/10.1103/PhysRevLett.103.083601}{Phys. Rev. Lett. \textbf{103}, 083601 (2009)}.

\bibitem {Hou_PRL2019}J. T. Hou and L. Liu, Strong Coupling between Microwave Photons and Nanomagnet Magnons, \href{https://link.aps.org/doi/10.1103/PhysRevLett.123.107702}{Phys. Rev. Lett. \textbf{123}, 107702 (2019)}.

\bibitem {Li_PRL2019}Y. Li, T. Polakovic, Y.-L. Wang, J. Xu, S. Lendinez, Z. Zhang, J. Ding, T. Khaire, H. Saglam, R. Divan, J. Pearson, W.-K. Kwok, Z. Xiao, V. Novosad, A. Hoffmann, and W. Zhang, Strong Coupling between Magnons and Microwave Photons in On-Chip Ferromagnet-Superconductor Thin-Film Devices, \href{https://link.aps.org/doi/10.1103/PhysRevLett.123.107701}{Phys. Rev. Lett. \textbf{123}, 107701 (2019)}.

\bibitem {Kajiwara2010}Y. Kajiwara, K. Harii, S. Takahashi, J. Ohe, K. Uchida, M. Mizuguchi, H. Umezawa, H. Kawai, K. Ando, K. Takanashi, S. Maekawa, and E. Saitoh, Transmission of electrical signals by spin-wave interconversion in a magnetic insulator, \href{https://doi.org/10.1038/nature08876}{Nature (London) \textbf{464}, 262 (2010)}.

\bibitem {Xiong2020}Y. Xiong, Y. Li, M. Hammami, R. Bidthanapally, J. Sklenar, X. Zhang, H. Qu, G. Srinivasan, J. Pearson, A. Hoffmann, V. Novosad, and W. Zhang, Probing magnon-magnon coupling in exchange coupled $\rm {Y_3Fe_5O_{12}}$/Permalloy bilayers with magneto-optical effects, \href{https://doi.org/10.1038/s41598-020-69364-6}{Sci. Rep. \textbf{10}, 12548 (2020)}.

\bibitem {PhysRevLett_Li2020}Y. Li, W. Cao, V. P. Amin, Z. Zhang, J. Gibbons, J. Sklenar, J. Pearson, P. M. Haney, M. D. Stiles, W. E. Bailey, V. Novosad, A. Hoffmann, and W. Zhang, Coherent Spin Pumping in a Strongly Coupled Magnon-Magnon Hybrid System,  \href{https://link.aps.org/doi/10.1103/PhysRevLett.124.117202}{Phys. Rev. Lett. \textbf{124}, 117202 (2020)}.

\bibitem {PhysRevLett_Klingler2018}S. Klingler, V. Amin, S. Gepr\"{a}gs, K. Ganzhorn, H. Maier-Flaig, M. Althammer, H. Huebl, R. Gross, R. D. McMichael, M. D. Stiles, S. T. B. Goennenwein, and M. Weiler, Spin-Torque Excitation of Perpendicular Standing Spin Waves in Coupled YIG/Co Heterostructures, \href{https://link.aps.org/doi/10.1103/PhysRevLett.120.127201}{Phys. Rev. Lett. \textbf{120}, 127201 (2018)}.

\bibitem {PhysRevLett_Chen2018}J. Chen, C. Liu, T. Liu, Y. Xiao, K. Xia, G. E. W. Bauer,M. Wu, and H. Yu, Strong Interlayer Magnon-Magnon Coupling in Magnetic Metal-Insulator Hybrid Nanostructures, \href{https://link.aps.org/doi/10.1103/PhysRevLett.120.217202}{Phys. Rev. Lett. \textbf{120}, 217202 (2018)}.

\bibitem {Qin2018}H. Qin, S. J. H\"{a}m\"{a}l\"{a}inen, and S. van Dijken, Exchange-torque-induced excitation of perpendicular standing spin waves in nanometer-thick $\rm {YIG}$ films, \href{https://doi.org/10.1038/s41598-018-23933-y}{ Sci. Rep. \textbf{8}, 5755 (2018)}.

\bibitem {Liu2018}C. Liu, J. Chen, T. Liu, F. Heimbach, H. Yu, Y. Xiao, J. Hu, M. Liu, H. Chang, T. Stueckler, S. Tu, Y. Zhang, Y. Zhang, P. Gao, Z. Liao, D. Yu, K. Xia, N. Lei, W. Zhao, and M. Wu, Long-distance propagation of short-wavelength spin waves, \href{https://doi.org/10.1038/s41467-018-03199-8}{Nat. Commun. \textbf{9}, 738 (2018)}.

\bibitem {PhysRevApplied_An2019}K. An, V. Bhat, M. Mruczkiewicz, C. Dubs, and D. Grundler, Optimization of spin-wave propagation with enhanced group velocities by exchange-coupled ferrimagnet-ferromagnet bilayers, \href{https://link.aps.org/doi/10.1103/PhysRevApplied.11.034065}{Phys. Rev. Applied \textbf{11}, 034065 (2019)}.

\bibitem {Kalinikos_1986}B. A. Kalinikos and A. N. Slavin, Theory of dipole-exchange spin wave spectrum for ferromagnetic films with mixed exchange boundary conditions \href{https://doi.org/10.1088/0022-3719/19/35/014}{J. Phys. C: Solid State Phys. \textbf{19}, 7013 (1986)}.

\bibitem {RADO_1959}G.T. Rado and J.R. Weertman, Spin-wave resonance in a ferromagnetic metal, \href{https://doi.org/10.1016/0022-3697(59)90233-1}{J. Phys. Chem. Solids \textbf{11}, 315 (1959)}.

\bibitem {RAME_1976}O. G. Ramer and C. H. Wilts, The Effects of Surface Layers on Spin-Wave Spectra in YIG Films, \href{http://onlinelibrary.wiley.com/doi/10.1002/pssb.2220730208}{Phys. Stat. Sol. b \textbf{73}, 443 (1976)}.

\bibitem {PUSZKARSKI_1976}H. Puszkarski, Theory of surface states in spin wave resonance, \href{http://www.sciencedirect.com/science/article/pii/0079681679900133}{Prog. Surf. Phys. \textbf{9}, 119 (1979)}.

\bibitem {Barnas_JMMM1991}J. Barna\'{s}, On the Hoffmann boundary conditions at the interface between two ferromagnets, \href{https://www.sciencedirect.com/science/article/abs/pii/030488539190145Z}{J. Magn. Magn. Mater. \textbf{102}, 319 (1991)}.

\bibitem {Mills_PRB1992}D. L. Mills, Spin waves in ultrathin exchange-coupled ferromagnetic multilayers: The boundary condition at the interface, \href{https://journals.aps.org/prb/abstract/10.1103/PhysRevB.45.13100}{Phys. Rev. B \textbf{45}, 13100 (1992)}.

\bibitem {Gui_PRL_2007}Y. S. Gui, N. Mecking, and C. M. Hu, Quantized Spin Excitations in a Ferromagnetic Microstrip from Microwave Photovoltage Measurements, \href{https://link.aps.org/doi/10.1103/PhysRevLett.98.217603}{Phys. Rev. Lett. \textbf{98}, 217603 (2007)}.

\bibitem {Kruglyak_JPCM2014}V. V. Kruglyak and O. Y. Gorobets and Y. I. Gorobets and A. N. Kuchko, Magnetization boundary conditions at a ferromagnetic interface of finite thickness, \href{https://iopscience.iop.org/article/10.1088/0953-8984/26/40/406001}{J. Phys.: Condens. Matter \textbf{26}, 406001 (2014)}.

\bibitem {Yin_PRL_2006}L. F. Yin, D. H. Wei, N. Lei, L. H. Zhou, C. S. Tian, G. S. Dong, X. F. Jin, L. P. Guo, Q. J. Jia, and R. Q. Wu, Magnetocrystalline Anisotropy in Permalloy Revisited, \href{https://journals.aps.org/prl/abstract/10.1103/PhysRevLett.97.067203}{Phys. Rev. Lett. \textbf{97}, 067203 (2006)}.

\bibitem {Stancil_2009}D. D. Stancil and A. Prabhakar, \emph{Spin Waves: Theory and Applications. Appendix A} (Springer, New York, 2009).

\bibitem {Wang_PRL_2020}H. Wang, J. Chen, T. Liu, J. Zhang, K. Baumgaertl, C. Guo, Y. Li, C. Liu, P. Che, S. Tu, S. Liu, P. Gao, X. Han, D. Yu, M. Wu, D. Grundler, and H. Yu, Chiral Spin-Wave Velocities Induced by All-Garnet Interfacial Dzyaloshinskii-Moriya Interaction in Ultrathin Yttrium Iron Garnet Films, \href{https://journals.aps.org/prl/cited-by/10.1103/PhysRevLett.124.027203}{Phys. Rev. Lett. \textbf{124}, 027203 (2020)}.

\bibitem {Rado_PR1954}G. T. Rado and J. R. Weertman, Observation of Exchange Interaction Effects in Ferromagnetics by Spin Wave Resonance, \href{https://journals.aps.org/pr/abstract/10.1103/PhysRev.94.1386}{Phys. Rev. \textbf{94}, 1386 (1954)}.

\bibitem {Puszkarski_2016}H. Puszkarski, Rado-Weertman Boundary Equation Revisited in Terms of the Free-Energy Density of a Thin Film, \href{http://przyrbwn.icm.edu.pl/APP/PDF/129/a129z6pr1.pdf}{Acta Phys. Pol. A \textbf{129}, RK.129.6.1 (2016)}.


\bibitem {Tahir_PRB2015}N. Tahir, R. Bali, R. Gieniusz, S. Mamica, J. Gollwitzer, T. Schneider, K. Lenz, K. Potzger, J. Lindner, M. Krawczyk, J. Fassbender, and A. Maziewski, Tailoring dynamic magnetic characteristics of Fe$_{60}$Al$_{40}$ films through ion irradiation, \href{https://journals.aps.org/prb/abstract/10.1103/PhysRevB.92.144429}{Phys. Rev. B \textbf{92}, 144429 (2015)}.

\bibitem {Klos_PRB2012}K\l{}os, J. W. and Kumar, D. and Romero-Vivas, J. and Fangohr, H. and Franchin, M. and Krawczyk, M. and Barman, A., Effect of magnetization pinning on the spectrum of spin waves in magnonic antidot waveguides, \href{https://journals.aps.org/prb/abstract/10.1103/PhysRevB.86.184433}{Phys. Rev. B \textbf{86}, 184433 (2012)}.


\bibitem {Navabi_PRAppl2017}A. Navabi, C. Chen, A. Barra, M. Yazdani, G. Yu, M. Montazeri, M. Aldosary, J. Li, K. Wong, Q. Hu, J. Shi, G. P. Carman, A. E. Sepulveda, P. K. Amiri, and K. L. Wang, Efficient excitation of high-frequency exchange-dominated spin waves in periodic ferromagnetic structures, \href{https://link.aps.org/doi/10.1103/PhysRevApplied.7.034027}{Phys. Rev. Applied \textbf{7}, 034027 (2017)}.

\bibitem {Cochran_PRB_1986}J. F. Cochran, B. Heinrich, and A. S. Arrott, Ferromagnetic resonance in a system composed of a ferromagnetic substrate and an exchange-coupled thin ferromagnetic overlayer, \href{https://journals.aps.org/prb/abstract/10.1103/PhysRevB.34.7788}{Phys. Rev. B \textbf{34}, 7788 (1986)}.

\bibitem {Magaraggia_PRB_2011}R. Magaraggia, K. Kennewell, M. Kostylev, R. L. Stamps, M. Ali, D. Greig, B. J. Hickey, and C. H. Marrows, Exchange anisotropy pinning of a standing spin-wave mode, \href{https://journals.aps.org/prb/references/10.1103/PhysRevB.83.054405}{Phys. Rev. B \textbf{83}, 054405 (2011)}.

\bibitem {Kostylev_JAP2014}M. Kostylev, Interface boundary conditions for dynamic magnetization and spin wave dynamics in a ferromagnetic layer with the interface Dzyaloshinskii-Moriya interaction, \href{https://aip.scitation.org/doi/10.1063/1.4883181}{J. Appl. Phys. \textbf{115}, 233902 (2014)}.

\bibitem {Vohl_PRB1989}M. Vohl, J. Barna\'{s}, and P. Gr\"{u}nberg, Effect of interlayer exchange coupling on spin-wave spectra in magnetic double layers: Theory and experiment, \href{https://doi.org/10.1103/PhysRevB.39.12003}{Phys. Rev. B \textbf{39}, 12003 (1989)}.

\bibitem {BARNAS_JMMM1989}J. Barna\'{s} and P. Gr\"{u}nberg, Spin waves in exchange-coupled epitaxial double-layers, \href{https://www.sciencedirect.com/science/article/abs/pii/0304885389901534}{J. Magn. Magn. Mater. \textbf{82}, 186 (1989)}.

\bibitem {Stamps_PRB1994}R. L. Stamps, Spin configurations and spin-wave excitations in exchange-coupled bilayers, \href{https://journals.aps.org/prb/abstract/10.1103/PhysRevB.49.339}{Phys. Rev. B \textbf{49}, 339 (1994)}.

\bibitem {Hillebrands_PRB1990}B. Hillebrands, Spin-wave calculations for multilayered structures, \href{https://journals.aps.org/prb/cited-by/10.1103/PhysRevB.41.530}{Phys. Rev. B \textbf{41}, 530 (1990)}.

\bibitem {Krawczyk_JPCM2003}M. Krawczyk, H. Puszkarski, J-C S. L\'{e}vy and D. Mercier, Exchange anisotropy pinning of a standing spin-wave mode, \href{https://doi.org/10.1088/0953-8984/15/17/303}{J. Phys.: Condens. Matter \textbf{15}, 2449 (2003)}.

\bibitem {Kruglyak_JPD2017}V. V. Kruglyak, C. S. Davies, V. S. Tkachenko, O. Y. Gorobets, Y. I. Gorobets and A. N. Kuchko, Formation of the band spectrum of spin waves in 1D magnonic crystals with different types of interfacial boundary conditions, \href{https://iopscience.iop.org/article/10.1088/1361-6463/aa536c}{J. Phys. D: Appl. Phys. \textbf{50}, 094003 (2017)}.

\bibitem {Vukadinovic_PRB2009}N. Vukadinovic, J. Ben Youssef, V. Castel, and M. Labrune, Magnetization dynamics in interlayer exchange-coupled in-plane/out-of-plane anisotropy bilayers, \href{https://journals.aps.org/prb/abstract/10.1103/PhysRevB.79.184405}{Phys. Rev. B \textbf{79}, 184405 (2009)}.

\bibitem {Balaz_PRB2015}Pavel Bal\'{a}\u{z} and J\'{o}zef Barna\'{s}, Magnetization dynamics in interlayer exchange-coupled in-plane/out-of-plane anisotropy bilayers, \href{https://journals.aps.org/prb/abstract/10.1103/PhysRevB.79.184405}{Phys. Rev. B \textbf{79}, 184405 (2009)}.

\bibitem {Hoffmann_1970jap}F. Hoffmann, A. Stankoff, and H. Pascard, Evidence for an exchange coupling at the interface between two ferromagnetic films, \href{https://doi.org/10.1063/1.1658798}{J. Appl. Phys. \textbf{41}, 1022 (1970)}.

\bibitem {Hoffmann_1970pss}F. Hoffmann, Dynamic pinning induced by nickel layers on permalloy films, \href{https://doi.org/10.1002/pssb.19700410237}{Phys. Stat. Sol. b \textbf{41}, 807 (1970)}.

\bibitem {Cochran PRB1992}J. F. Cochran and B. Heinrich, Boundary conditions for exchange-coupled magnetic slabs, \href{https://journals.aps.org/prb/abstract/10.1103/PhysRevB.45.13096}{Phys. Rev. B \textbf{45}, 13096 (1992)}.

\bibitem {Li_AFM_2016}S. Li, Q. Li, J. Xu, S. Yan, G.-X. Miao, S. Kang, Y. Dai, J. Jiao, and Y. L\"{u}, Tunable optical mode ferromagnetic resonance in $\rm {FeCoB/Ru/FeCoB}$ synthetic antiferromagnetic trilayers under uniaxial magnetic anisotropy,  \href{https://doi.org/10.1002/adfm.201600122}{Adv. Funct. Mater. \textbf{26}, 3738 (2016)}.

\bibitem {Li_AMI_2018}S. Li, G.-X. Miao, D. Cao, Q. Li, J. Xu, Z. Wen, Y. Dai, S. Yan, and Y. L\"{u}, Stress-enhanced interlayer exchange coupling and optical-mode FMR frequency in self-bias FeCoB/Ru/FeCoB trilayers,  \href{https://doi.org/10.1021/acsami.7b19684}{ACS Appl. Mater. Interfaces  \textbf{10}, 8853 (2018)}.

\bibitem {Wang_APL2018}W. Wang, P. Li, C. Cao, F. Liu, R. Tang, G. Chai, and C. Jiang, Temperature dependence of interlayer exchange coupling and gilbert damping in synthetic antiferromagnetic trilayers investigated using broadband ferromagnetic resonance,  \href{https://doi.org/10.1063/1.5040666}{Appl. Phys. Lett. \textbf{113}, 042401 (2018)}.

\bibitem {Zhang_PRL1994}Z. Zhang, L. Zhou, P. E. Wigen, and K. Ounadjela, Using Ferromagnetic Resonance as a Sensitive Method to Study Temperature Dependence of Interlayer Exchange Coupling, \href{https://journals.aps.org/prl/abstract/10.1103/PhysRevLett.73.336}{Phys. Rev. Lett. \textbf{73}, 336 (1994)}.

\bibitem {Khodadadi_PRAppl2017}B. Khodadadi, J. B. Mohammadi, J. M. Jones, A. Srivastava, C. Mewes, T. Mewes, and C. Kaiser, Interlayer Exchange Coupling in Asymmetric Co\-Fe/Ru/Co\-Fe Trilayers Investigated with Broadband Temperature-Dependent Ferromagnetic Resonance, \href{https://journals.aps.org/prapplied/abstract/10.1103/PhysRevApplied.8.014024}{Phys. Rev. Applied \textbf{8}, 014024 (2017)}.

\bibitem {Heinrich_PRL2003}B. Heinrich, Y. Tserkovnyak, G. Woltersdorf, A. Brataas, R. Urban, and G. E. W. Bauer, Dynamic Exchange Coupling in Magnetic Bilayers, \href{https://link.aps.org/doi/10.1103/PhysRevLett.90.187601}{Phys. Rev. Lett. \textbf{90}, 187601 (2003)}.

\bibitem {Tserkovnyak_RMP2005}Y. Tserkovnyak, A. Brataas, G. E. W. Bauer, and B. I. Halperin, Nonlocal magnetization dynamics in ferromagnetic heterostructures, \href{https://link.aps.org/doi/10.1103/RevModPhys.77.1375}{Rev. Mod. Phys. \textbf{77}, 1375 (2005)}.

\bibitem {ZHANG201879}S. Zhang, J. Rong, H. Wang, D. Wang, and L. Zhang, Spin-wave resonance frequency in ferromagnetic thin film with interlayer exchange coupling and surface anisotropy, \href{https://doi.org/10.1016/j.susc.2017.09.005}{Surf. Sci. \textbf{667}, 79 (2018)}.

\bibitem {MAILIAN2019484}M. Mailian, O. Gorobets, Y. Gorobets, M. Zelent, and M. Krawczyk, Exchange spin waves transmission through the interface between two antiferromagnetically coupled ferromagnetic media, \href{https://doi.org/10.1016/j.jmmm.2019.02.062}{J. Magn. Magn. Mater.  \textbf{484}, 484 (2018)}.

\bibitem {MAILIAN2018}M. Milian, O. Gorobets, Y. Gorobets, M. Zelent, and M. Krawczyk, Control of the spin wave phase in transmission through the ultrathin interface between exchange coupled ferromagnetic materials, \href{http://doi.org/10.12693/APhysPolA.133.480}{Acta Phys. Pol. A  \textbf{133}, 480 (2018)}.

\bibitem {Verba_2020}R. Verba, V. Tiberkevich, and A. Slavin, Spin-wave transmission through an internal boundary: Beyond the scalar approximation, \href{https://link.aps.org/doi/10.1103/PhysRevB.101.144430}{Phys. Rev. B \textbf{101}, 144430 (2020)}.

\bibitem {Herring_PR1951}C. Herring and C. Kittel, On the Theory of Spin Waves in Ferromagnetic Media, \href{https://journals.aps.org/pr/abstract/10.1103/PhysRev.81.869}{Phys. Rev. \textbf{81}, 869 (1951)}.

\bibitem {Demokritov_PR2001}S.O. Demokritov and B. Hillebrands and A.N. Slavin, Brillouin light scattering studies of confined spin waves: linear and nonlinear confinement, \href{http://www.sciencedirect.com/science/article/pii/S0370157300001162}{Phys. Rep. \textbf{348}, 441 (2001)}.

\bibitem {Akhiezer1968SpinWaves}A. I. Akhiezer, V. G. Bar'Yakhtar, and S. V. Peletminskii, \emph{Spin Waves} (Amsterdam: North-Holland, 1968).

\bibitem {Arne_2014}A. Vansteenkiste, J. Leliaert, M. Dvornik, M. Helsen, F. GarciaSanchez, and B. Van Waeyenberge, The design and verification of mumax3,  \href{https://doi.org/10.1063/1.4899186}{AIP Adv. \textbf{4}, 107133 (2014)}.

\bibitem {Sushruth_PRR2020}M. Sushruth, M. Grassi, K. Ait-Oukaci, D. Stoeffler, Y. Henry, D. Lacour, M. Hehn, U. Bhaskar, M. Bailleul, T. Devolder, and J.-P. Adam, Electrical spectroscopy of forward volume spin waves in perpendicularly magnetized materials, \href{https://journals.aps.org/prresearch/abstract/10.1103/PhysRevResearch.2.043203}{Phys. Rev. Research \textbf{2}, 043203 (2020)}.

\bibitem {GELLER195730}S. Geller and M. Gilleo, The crystal structure and ferrimagnetism of yttrium-iron garnet, $\rm {Y_3Fe_2(FeO_4)_3}$, \href{https://doi.org/10.1016/0022-3697(57)90044-6}{J. Phys. Chem. Solid. \textbf{3}, 30 (1957)}.

\bibitem {Fan_PRAppl2020}Y. Fan, P. Quarterman, J. Finley, J. Han, P. Zhang, J. T. Hou, M. D. Stiles, A. J. Grutter, and L. Liu, Manipulation of coupling and magnon transport in magnetic metal-insulator hybrid structures, \href{https://link.aps.org/doi/10.1103/PhysRevApplied.13.061002}{Phys. Rev. Applied \textbf{13}, 061002 (2020)}.

\bibitem {Chun_JAP2004}Y. S. Chun and K. M. Krishnana, Interlayer perpendicular domain coupling between thin Fe films and garnet single-crystal underlayers, \href{https://aip.scitation.org/doi/10.1063/1.1689909}{J. Appl. Phys. \textbf{95}, 6858 (2004)}.

\bibitem {Nahrwold_JAP2010}G. Nahrwold, J. M. Scholtyssek, S. Motl-Ziegler, O. Albrecht, U. Merkt, and G. Meier, Structural, magnetic, and transport properties of Permalloy for spintronic experiments, \href{https://aip.scitation.org/doi/citedby/10.1063/1.3431384}{J. Appl. Phys. \textbf{108},  013907 (2010)}.

\bibitem {Mallmann_SSP2013}E. J. J. Mallmann, A. S. B. Sombra, J. C. Goes, and P. B. A. Fechine, Yttrium Iron Garnet: Properties and Applications Review, \href{https://www.scientific.net/SSP.202.65}{Solid State Phenom. \textbf{202},  65 (2013)}.

\bibitem {Poimanov_PRB2018}V. D. Poimanov, A. N. Kuchko, and V. V. Kruglyak, Magnetic interfaces as sources of coherent spin waves, \href{https://link.aps.org/doi/10.1103/PhysRevB.98.104418}{Phys. Rev. B \textbf{98}, 104418 (2018)}.



\bibitem {VAYHINGER1988307}K. Vayhinger and H. Kronm\"{u}ller, Spin wave theory of exchange coupled ferromagnetic multilayers, \href{https://doi.org/10.1016/0304-8853(88)90227-2}{J. Magn. Magn. Mater. \textbf{72}, 307 (1988)}.


\bibitem {Morrison_JAP2015}C. Morrison, J. J. Miles, T. N. Anh Nguyen, Y. Fang, R. K. Dumas, J. {\AA}kerman, and T. Thomson, Exchange coupling in hybrid anisotropy magnetic multilayers quantified by vector magnetometry, \href{https://doi.org/10.1063/1.4917336}{J. Appl. Phys. \textbf{117}, 17B526 (2015)}.

\end{thebibliography}
\end{document}